\documentclass[12pt]{article}

\usepackage{fullpage,amsfonts,graphicx,amsmath,epsfig}

\newcommand{\T}{\mathbb{T}}
\newcommand{\Z}{\mathbb{Z}}
\newcommand{\D}{\mathbb{D}}

\newtheorem{theorem}{Theorem}

\newcommand{\proof}{\noindent {\bf Proof:} }
\newcommand{\qed}{\hfill {\bf QED}}

\begin{document}

\title{Weak chimeras in minimal networks of coupled phase oscillators}

\author{Peter Ashwin\\
Centre for Systems, Dynamics and Control,\\
Harrison Building,\\
University of Exeter,  Exeter EX4 4QF, UK
\and
Oleksandr Burylko\\
Institute of Mathematics, National Academy of Sciences\\
01601 Kyiv, Ukraine
}
\maketitle

\begin{abstract}
We suggest a definition for a type of chimera state that appears in networks of indistinguishable phase oscillators. Defining a ``weak chimera'' as a type of invariant set showing partial frequency synchronization, we show that this means they cannot appear in phase oscillator networks that are either globally coupled or too small. We exhibit various networks of four, six and ten indistinguishable oscillators where weak chimeras exist with various dynamics and stabilities. We examine the role of Kuramoto-Sakaguchi coupling in giving degenerate (neutrally stable) families of weak chimera states in these example networks.
\end{abstract}

\section{Introduction}

{\bf
Coupled oscillator systems are a rich source of examples of high dimensional dynamical behaviour as well as a class of systems that can be used to understand a range of emergent dynamical phenomena. One of these phenomena, where there is apparent coexistence of coherent and incoherent behaviour, has been called a chimera state.  This paper proposes a definition of a ``weak chimera'' for finite networks of coupled indistinguishable phase oscillators. This definition is relatively easily checkable from the dynamics and allows us to prove existence as well as investigating stability and bifurcations of weak chimeras in small networks. Although chimeras in many high dimensional systems are not weak chimeras in the sense we define here, we suggest that weak chimeras may be responsible for organizing the dynamics of more general chimera states.
}

Kuramoto's model for globally coupled phase oscillators
$$
\dot{\theta}_i= \frac{d}{dt}\theta_i =\omega_i - \frac{K}{N} \sum_{j=1}^{N} \sin (\theta_i-\theta_j)
$$
with $\theta_i\in[0,2\pi)$, $\omega_i$ and $K$ constant \cite{kuramoto} has been used for many years as a prototype of an oscillator system where sufficiently strong $K>0$ will result in synchrony. Dynamically more complex solutions include partial synchrony or clustering. For phase oscillator networks that are not globally coupled, some intriguing solutions were first noted by Kuramoto and Battogtokh \cite{kuramoto-2003} and named ``chimera states'' by Abrams and Strogatz \cite{abrams-strogatz-2004,abrams-strogatz-2006}. In these states, the oscillators split into two (or more) regions one of which is coherent while the other is incoherent, in some sense. 

A number of authors have studied chimera states in a wide range of contexts, for example \cite{laing-2010,martens-2010,omelchenko-etal-2010,sieber-etal-2014,panaggio-abrams-2014} and it seems that rather than being exceptional they are, in some sense, prevalent. Most work on chimeras has however not attempted to make a rigorous definition of chimera state that can easily be applied to small systems. For instance, \cite{abrams-strogatz-2006} state that ``For certain choices of parameters and initial conditions, the array would split into two domains: one composed of coherent, phase-locked oscillators, coexisting with another composed of incoherent, drifting oscillators'' but in particular, the words ``domain'', ``incoherence'' and ``drifting'' need careful interpretation before they can be applied to small systems.

The paper is organized as follows: In section~\ref{sec:indist} we consider some basic dynamical properties of networks of indistinguishable phase oscillators and propose a definition of weak chimera state for these systems. Section~\ref{sec:reduction} gives a basic result on the non-existence of weak chimera states for globally coupled phase oscillator networks, and then looks at minimal networks of four, six and ten phase oscillators where a modular structure allows us to prove there are weak chimera attractors. The detailed dynamics of these examples are at least quasiperiodic but may in principle be much more complex - for example, the ten oscillator example has a weak chimera attractor that is an attracting heteroclinic network. Section~\ref{sec:nonmod} discusses an example of a non-modular network (a ring of six oscillators with nearest and next-nearest neighbour coupling) where one can find attracting weak chimera states and investigate the bifurcations that create them. Finally, Section~\ref{sec:discuss} discusses some of the consequences and limitations of these results. In particular we note that the special case of Kuramoto-Sakaguchi coupling (often considered for phase oscillator chimera examples) has families of weak chimeras with degenerate stability. We suggest that this may be related to the fact that chimeras appear to be transients in simulations of small networks \cite{wolfrum-omelchenko-2011}.

\section{Weak chimeras in networks of indistinguishable phase oscillators}
\label{sec:indist}

Consider a system of $N$ coupled phase oscillators described as an ODE on the torus $(\theta_1,\ldots,\theta_N)\in\T^N=[0,2\pi)^N$:
\begin{equation}
\dot{\theta}_i =\omega_i+\sum_{j=1}^{N} K_{ij} g(\theta_i-\theta_j)
\label{eq:model}
\end{equation}
where $K_{ij}$ is the strength of coupling, $\omega_i$ is the natural frequency of the $i$th oscillator and $g(\phi)$ is a smooth $2\pi$-periodic coupling function. The phase oscillators are {\em identical} if $\omega_i=\omega$, and if we are interested in phase differences, we can set $\omega=0$ without loss of generality. We consider Hansel-Mato-Meunier coupling \cite{hansel-mato-meunier-1993,ashwin-burylko-maistrenko-2008} with parameters $\alpha$ and $r$:
\begin{equation}
g(\phi)=-\sin(\phi-\alpha)+r\sin (2\phi)
\label{eq:hmmcoupling}
\end{equation}
which reduces to Kuramoto-Sakaguchi coupling \cite{sakaguchi_kuramoto_86} for $r=0$.

We say the oscillators are {\em indistinguishable} if the oscillators are identical and interchangeable in the sense that they have the same number and strength of inputs \cite[Def~3.2]{ashwin-swift-1992}. Let $S_N$ denote the permutation group acting on the $N$ oscillator phases. Equivalent ways of expressing this are:
\begin{itemize}
\item[(a)] Only one equation is needed to specify the system, up to permutation of indices.
\item[(b)] There are $N$ permutations $\sigma_i \in S_N$ with $\sigma_i(i)=i$ for $i=1,\cdots,N$ such that the matrix $K_{ij}$ satisfies
$$
K_{ij}=k_{\sigma_i(j)}
$$
for some vector $k_i$ and for all $i\neq j$; namely the matrix is a permutation of a vector of coupling strengths.
\item[(c)] The system is invariant under a permutation symmetry group that acts transitively on the set of $N$ oscillators.
\end{itemize}
Figures~\ref{fig:networks} and \ref{fig:modular_graphs} illustrate some examples of small networks where the oscillators are indistinguishable. We say oscillators $i$ and $j$ on a trajectory of the system (\ref{eq:model}) are {\em frequency synchronized} if 
$$
\Omega_{ij}:=\lim_{T\rightarrow \infty} \frac{1}{T}[\theta_i(T)-\theta_j(T)]=0
$$
where we choose continuous representatives for $\theta_{i}(t),\theta_j(t)$ (N.B. is not necessary for the oscillators to have well-defined frequencies for the system to be frequency synchronized \cite{karabacak_ashwin_10}).

\begin{quote}
{\bf We say $A\subset\T^N$ is a {\em weak chimera} state for a coupled phase oscillator system if it is a connected chain-recurrent \cite{franke-selgrade-1976} flow-invariant set such that on each trajectory within $A$ there are $i,j$ and $k$ such that $\Omega_{ij}\neq 0$ and $\Omega_{ik}=0$.}
\end{quote}

We do not place any restriction on the dynamical behaviour or stability of $A$: if it is of saddle type or has neutral stability the behaviour would only be visible as a transient for typical initial conditions \cite{wolfrum-omelchenko-2011}. If $A$ is the $\omega$-limit of some initial condition then $A$ is {\em connected} and {\em chain-recurrent} \cite{franke-selgrade-1976}. Hence we include these as necessary conditions for the dynamics of $A$ to be visible in the long-term behaviour of a single trajectory.

Due to the drift of the incoherent region, the chimera states of \cite{kuramoto-2003,abrams-strogatz-2004,abrams-strogatz-2006} for large $N$ are in fact not weak chimeras. However as we discuss in Section~\ref{sec:discuss}, unstable weak chimeras may play an important role in organizing such chimeras in coupled phase oscillator networks, just as unstable periodic orbits play an important role in organizing chaotic dynamics.

There is an element of surprise in the definition of weak chimera: one might expect systems of indistinguishable phase oscillators to always have frequency synchrony, but we will see that this is not the case for many networks. However, for some types of network there are obstructions to the existence of weak chimera states. For other networks we find parameters with {\em attracting} weak chimera states where the following hold:
\begin{itemize}
\item[(a)] there are at least four oscillators
\item[(b)] at least two different coupling strengths are present in the network
\item[(c)] there are at least two Fourier components in the coupling function (i.e., if coupling function is (\ref{eq:hmmcoupling}), then $r\neq 0$).
\end{itemize}
Note that (b) necessarily implies (a) for indistinguishable phase oscillators. Examples in the literature suggest that (c) is not necessary for existence of weak chimera states but we believe it may be for weak chimera states to be bistable with full synchrony.

\section{Indistinguishable phase oscillators and weak chimera states}
\label{sec:reduction}

For global (equal and all-to-all) coupling we write $K_{ij}=K$ and the system has full permutation symmetry $S_N$ \cite{ashwin-swift-1992}. As a consequence there is an invariant subspace corresponding to $\theta_i=\theta_j$ (modulo $2\pi$) for any $i\neq j$. The presence of $(N-1)!$ of these codimension one invariant subspaces implies that there will be a permutation of the oscillators $k(j)$ such that
\begin{equation}
\theta_{k(1)} \leq \theta_{k(2)}\leq \cdots \leq \theta_{k(N)} \leq \theta_{k(1)}+2\pi
\label{eq:canonical}
\end{equation}
is satisfied along the trajectory. This can be used to show a result (already effectively stated in \cite{abrams-strogatz-2006}):

\begin{theorem}{\bf \cite[Lemma~5.3]{ashwin-swift-1992}}\label{prop:alltoall}
For global coupling of $N$ identical phase oscillators with $K_{ij}=K$ and any $g(\phi)$, all trajectories of (\ref{eq:model}) are frequency synchronized. Hence no weak chimera states are possible in such a system.
\end{theorem}

This result does not generalise to more general oscillators with global coupling as higher dimensional systems do not necessarily satisfy (\ref{eq:canonical}). Indeed, \cite{schmidt-etal-2014,sethia-sen-2014} find chimera states in globally coupled networks with two-dimensional oscillators.

\begin{figure}%
\begin{center}
\includegraphics[width=13cm]{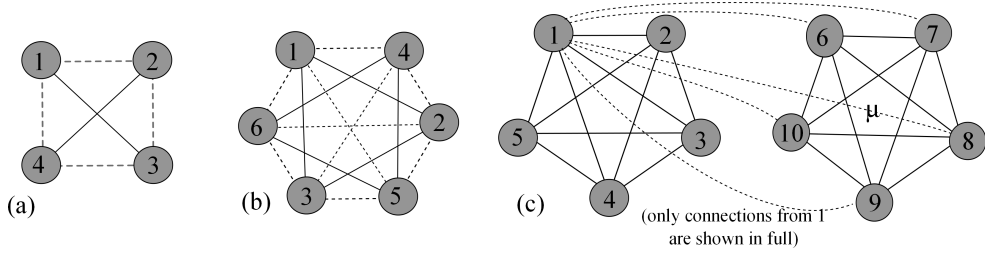}
\end{center}
\caption{Example networks of (a) four, (b) six and (c) ten indistinguishable oscillators that permit robust weak chimera states. The solid line indicates bidirectional coupling with strength $1$ while the dashed line indicates bidirectional coupling with strength $\epsilon$ (for clarity in (c), only oscillator one is shown with its full set of connections). Each the networks has a modular structure, i.e. they decouple into a number of smaller networks for $\epsilon=0$.}%
\label{fig:networks}%
\end{figure}

In the remainder of this section we show that weak coupling between two subnetworks (or modules) can give rise to weak chimera states; in particular for the networks shown in Figure~\ref{fig:networks}.

\subsection{Four oscillator example: stable weak chimera with in-phase and anti-phase groups}
\label{sec:fourcell}

Consider the system (\ref{eq:model},\ref{eq:hmmcoupling}) for $N=4$ with coupling as in Figure~\ref{fig:networks}(a) and coupling strengths $K_{ij}\in\{1,\epsilon\}$. This means that (\ref{eq:model}) can be written as
\begin{eqnarray}
\dot{\theta}_1 &=& \omega + (g(\theta_1-\theta_3)+g(0))+\epsilon (g(\theta_1-\theta_2)+g(\theta_1-\theta_4))\nonumber\\
\dot{\theta}_2 &=& \omega + (g(\theta_2-\theta_4)+g(0))+\epsilon (g(\theta_2-\theta_3)+g(\theta_2-\theta_1))\nonumber\\
\dot{\theta}_3 &=& \omega + (g(\theta_3-\theta_1)+g(0))+\epsilon (g(\theta_3-\theta_2)+g(\theta_3-\theta_4))\label{eq:model4osc}\\
\dot{\theta}_4 &=& \omega + (g(\theta_4-\theta_2)+g(0))+\epsilon (g(\theta_4-\theta_1)+g(\theta_4-\theta_3)).\nonumber
\end{eqnarray}

\begin{theorem}
\label{thm:fourosc}
There is an open set of $(r,\alpha)$ such that the four-oscillator system (\ref{eq:model4osc},\ref{eq:hmmcoupling}) has an attracting weak chimera state for $\epsilon=0$ that persists for all $\epsilon$ with $|\epsilon|$ sufficiently small.
\end{theorem}

\proof
We write $\phi_1=\theta_1-\theta_3$, $\phi_2=\theta_2-\theta_4$, $\phi_3=\theta_1-\theta_2$ 
and $g_{ij}=g(\theta_i-\theta_j)$ so that (\ref{eq:model4osc}) becomes
\begin{eqnarray*}
\dot{\phi}_1 &=& g_{13}- g_{31}+\epsilon(g_{12}+g_{14}-g_{32}-g_{34})\\ 
\dot{\phi}_2 &=& g_{24}- g_{42}+\epsilon(g_{21}+g_{23}-g_{41}-g_{43})\\ 
\dot{\phi}_3 &=& g_{13}- g_{24}+\epsilon(g_{12}+g_{14}-g_{21}-g_{23}).
\end{eqnarray*}
If we write $g(\phi)=(p(\phi)+q(\phi))/2$ where $p$ is even and $q$ is odd then we have
\begin{eqnarray}
\dot{\phi}_1 &=&  q(\phi_1) + 
\epsilon (g(\phi_3) + g(\phi_3+\phi_2)- g(-\phi_1+\phi_3) - g(\phi_2+ \phi_3 -\phi_1)) \nonumber\\
\dot{\phi}_2 &=& q(\phi_2) + 
\epsilon (g(-\phi_3) + g(\phi_1-\phi_3) - g(-\phi_2-\phi_3) - g(\phi_1-\phi_2-\phi_3)) \label{eq:fouroscnew}\\
\dot{\phi}_3 &=&  g(\phi_1) -  g(\phi_2) + 
\epsilon (g(\phi_3) + g(\phi_3+\phi_2) - g(-\phi_3) - g(\phi_1 -\phi_3))\nonumber
\end{eqnarray}
Now consider the case $\epsilon=0$ and $\phi=\phi_i$ with $i=1,2$: these satisfy
$\dot{\phi} = q(\phi)$ where 
$$
q(\phi)=g(\phi)-g(-\phi)= -2\sin \phi\cos\alpha + 2r\sin(2\phi)=2\sin \phi\left(-\cos\alpha+2r\cos \phi\right),
$$
which for $(r,\alpha)$ in the region of bistability of in-phase and antiphase solutions (resp. $\phi=0$ and $\phi=\pi$) with $q(\phi)=0$. Note that $q'(0)=-2\cos\alpha+4r$ and $q'(\pi)=2\cos\alpha+4r$, so there is bistability when $q'(0)<0$ and $q'(\pi)<0$. This is the case if $-\cos \alpha+2r<0$ and $\cos \alpha+ 2r<0$, i.e. when $r<-(\cos \alpha)/2$ and $r < ( \cos \alpha)/2$. This can be satisfied in the region of $(r,\alpha)$ where
\begin{equation}\label{eq:bistable_rt}
r < \min\{\cos \alpha, -\cos\alpha\}/2=-|\cos\alpha|/2.
\end{equation}
Consider an initial condition $(\phi_1,\phi_2,\phi_3)=(0,\pi,\xi)$. For $\epsilon=0$, this initial condition lies on the periodic orbit $(\phi_1(t),\phi_2(t),\phi_3(t))=(0,\pi,\Omega t+\xi)$ where $\Omega:=g(0)-g(\pi)= 2 \sin \alpha$ independent of $r$. This periodic orbit is a compact recurrent invariant set that is not frequency synchronized as long as $\alpha\neq
k\pi$, $k\in {\mathbb Z}$.

This periodic orbit is stable with Floquet exponents given by $0$, $q'(0)2\pi/\Omega$ and $q'(\pi)2\pi/\Omega$. Finally, hyperbolicity of the linearly stable periodicity implies unique continuation of this stable periodic orbit under small perturbations of parameters - in particular for any $(r,\alpha)$ satisfying (\ref{eq:bistable_rt}) and $\alpha\neq k\pi$ there is an $\epsilon_0(r,\alpha)$ such that there is persistence of this weak chimera state for all $\epsilon$ where  $|\epsilon|<\epsilon_0(r,\alpha)$.
\qed

~

We do not give upper bounds on $\epsilon_0(r,\alpha)$ except to note that $\epsilon_0\searrow 0$ on any path where $r+|\cos\alpha|/2\nearrow 0$. From Theorem~\ref{prop:alltoall}, the weak chimera state must disappear for $\epsilon=1$, hence $\epsilon_0(r,\alpha)<1$. This weak chimera is degenerate at $r=0$ as there is no bistability of in-phase and antiphase synchrony in this case. The curve $r=-\frac{1}{2}\cos\alpha$ corresponds to a subcritical pitchfork bifurcation in the invariant plane $\phi_1=0$ of the stable cycle with coordinate $\phi_2=\pi$ and two saddle periodic orbits with coordinates $\phi_2=\pm\arccos(\cos\alpha/(2 r))$ for $\epsilon=0$. The curve $r=\frac{1}{2}\cos\alpha$ corresponds to a subcritical pitchfork bifurcation in the invariant plane $\phi_2=\pi$ of the same stable cycle with $\phi_1=0$ and two saddle cycles with $\phi_1=\pm\arccos(\cos\alpha/(2 r))$.

The system Figure~\ref{fig:networks}(a) can be generalized by specifying coupling $\epsilon_1$ from $\theta_i$ to $\theta_{i+1}$ and with coupling $\epsilon_2$ from $\theta_i$ to $\theta_{i-1}$ ($\Z_4$ symmetry). Theorem~\ref{thm:fourosc} can be generalised in this case as follows: for $\epsilon_1=\epsilon_2=\epsilon$, the plane $\phi_1=0$ is invariant and there is a weak chimera with $\phi_1=0$. For $\epsilon_1\ne\epsilon_2$ the plane $\phi_1=0$ is no longer invariant but can still be shown to contain a weak chimera.

\subsection{A six oscillator example: stable weak chimera with in-phase and splay-phase groups}

\begin{figure}%
\begin{center}
\includegraphics[width=12cm]{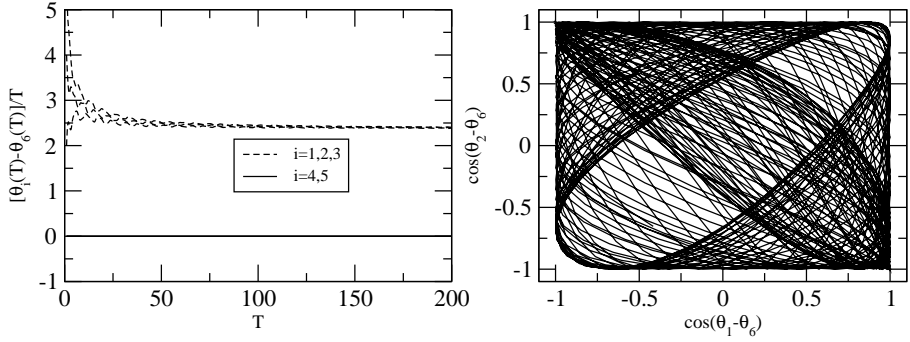}%
\end{center}
\caption{Example of a weak chimera attractor in the six oscillator system (\ref{eq:threeplusthree},\ref{eq:hmmcoupling}) as in Figure~\ref{fig:networks}(b). The oscillators $i=1,2,3$ limit to an approximately splay phase state while the oscillators $i=4,5,6$ limit to in-phase. The left panel shows convergence of $[\theta_i(T)-\theta_{6}(T)]/T$ towards well-defined frequency differences $\Omega_{i,6}$ such that $\Omega_{1,6}=\Omega_{2,6}=\Omega_{3,6}\neq 0$ and $\Omega_{4,6}=\Omega_{5,6}=0$. The right panel illustrates that the dynamics of the phase differences relative to the $6$th oscillator is quasiperiodic.
}
\label{fig:threeplusthree}%
\end{figure}

Consider the system with $N=6$ in Figure~\ref{fig:networks}(b) where
\begin{equation}
\dot{\theta}_{i+3j} = \omega + \sum_{k=1}^{3} \left[g(\theta_{i+3j}-\theta_{k+3j})+\epsilon g(\theta_{i+3j}-\theta_{k+3j+3})\right]
\label{eq:threeplusthree}
\end{equation}
with $i=1,\ldots,3$, $j=0,1$ and all subscripts are taken modulo 6. For (\ref{eq:hmmcoupling}) with $r=-0.15$, $\alpha=-1.7$ and $\epsilon=0.1$ there are chimera states where three of the oscillators are in-phase and the other three are close to a splay-phase (rotating wave, $\Z_3$) periodic orbit; see Figure~\ref{fig:threeplusthree}. 

One can see this as follows: for $\epsilon=0$ the systems splits into two groups of $N=3$ oscillators with all-to-all coupling and bistability of in-phase and splay-phase (anti-phase/rotating wave) solutions \cite[Fig 1]{ashwin-burylko-maistrenko-2008}. These solutions have distinct frequencies, so for $0<\epsilon\ll 1$ the system has attracting weak chimeras that are robust to small changes in the parameters.

\subsection{A ten oscillator example: stable weak chimera with in-phase and heteroclinic cycle groups}

\begin{figure}%
\begin{center}
\includegraphics[width=12cm]{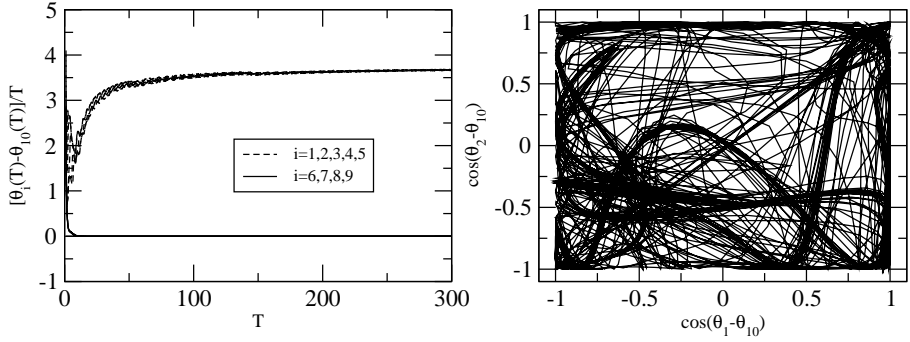}
\end{center}
\caption{Example of a weak chimera attractor in the system of ten oscillators (\ref{eq:tenosc},\ref{eq:tenosccoup}) in Figure~\ref{fig:networks}(c) where one group of five undergoes heteroclinic switching; see text for details. The oscillators $i=1,2,3,4,5$ approach heteroclinic cycle while oscillators $i=6,7,8,9,10$ are in-phase at a different frequency. The left panel shows convergence of $[\theta_i(T)-\theta_{10}(T)]/T$ towards well-defined frequency differences $\Omega_{i,10}$ such that $\Omega_{i,10}\neq 0$ for $i=1,2,3,4,5$ while $\Omega_{i,10}=0$ for the remaining group. The right panel illustrates that the dynamics of the phase differences relative to the $10$th oscillator is not simply periodic or quasiperiodic but switches between a number of saddle periodic orbits. As time progresses, the time spent near a periodic orbit gets progressively longer and longer.
}%
\label{fig:10oscchimera}%
\end{figure}

Consider the network Figure~\ref{fig:networks}(c) consisting of two groups of all-to-all coupled five oscillators and weak coupling between the groups, i.e.
\begin{equation}
\dot{\theta}_{i+5j} = \omega + \sum_{k=1}^{5} \left[g(\theta_{i+5j}-\theta_{k+5j})+\epsilon g(\theta_{i+5j}-\theta_{k+5j+5})\right]
\label{eq:tenosc}
\end{equation}
where $i=1,\ldots,5$, $j=0,1$ and all subscripts are taken modulo $10$. We choose
\begin{equation}
g(\phi)=-\sin(\phi-\alpha)+r\sin(2\phi-\beta)
\label{eq:tenosccoup}
\end{equation}
with $r=0.2$, $\alpha=4.67398$, $\beta=4.51239$, $\omega=0.1$
and $\epsilon=0.1$ such that there is a weak chimera where one group is in-phase while the other approaches a stable heteroclinic attractor; see Figure~\ref{fig:10oscchimera}. For $\epsilon=0$ where the two networks decouple, each is multistable with two attractors; in-phase synchrony and a heteroclinic network between 30 saddle periodic orbits are attractors. In the absence of noise, there will be switching between the saddle periodic orbits that progressively slows down; see \cite{ashwin_gabor_07} for a more detailed description of the heteroclinic network attractor,

\subsection{Weak chimera states and modular networks}
\label{sec:modular}

\begin{figure}%
\begin{center}
\includegraphics[width=10cm]{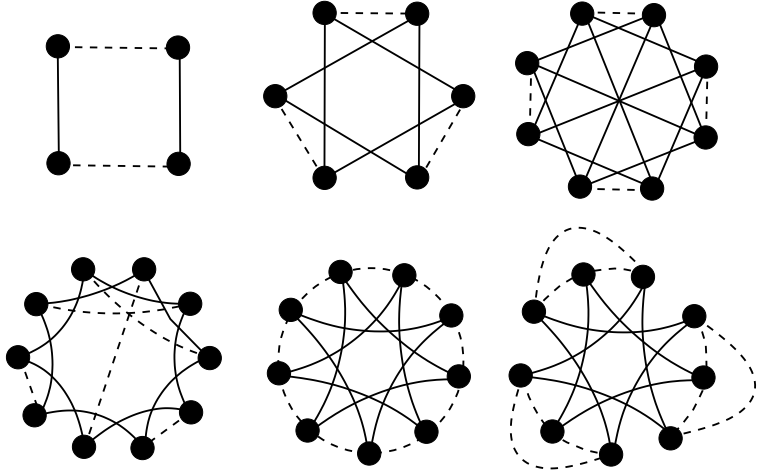}
\end{center}
\caption{More examples of indistinguishable oscillator networks with modular structure: each of these decouples into more than one identical networks on setting the coupling on the dashed lines to zero.}%
\label{fig:modular_graphs}%
\end{figure}

One can generalize the previous examples to networks of indistinguishable phase oscillators with modular structure. More precisely, suppose we have a system of $n=mk$ oscillators, $\theta\in \T^{m\times k}$, with $m>1$ and $k>1$ are integers, governed by
\begin{equation}\label{eq:modular_net}
\dot{\theta}_{ij} = \omega + \sum_{q=1}^{k}\left[K_{i j,i q}g(\theta_{ij}-\theta_{iq})  +\epsilon K_{i j,p q}\sum_{p=1,p\neq i}^{m} g(\theta_{ij}-\theta_{pq})\right]
\end{equation}
where $i=1,\ldots,m$, $j=1,\ldots,k$, $K_{i j, p q}\in\{0,1\}$ and $g$ is a smooth period coupling function (there will be constraints on $K_{i j, p q}$ for the oscillators to be indistinguishable). In such a case we say the system splits into $m$ modules of $k$ oscillators. The network decouples in the case $\epsilon=0$ into $m$ uncoupled but identical modules (networks) of $k$ phase oscillators. Each module is governed by the following equations for $\theta\in \T^{k}$, for some $L_{j q}\in\{0,1\}$:
\begin{equation}\label{eq:module}
\dot{\theta}_{j} = \omega + \sum_{q=1}^{k} L_{j q} g(\theta_{j}-\theta_{q}).
\end{equation}
If the module is multistable one can obtain sufficient conditions for the existence weak chimera states for $\epsilon>0$. Even for the case of modules with a hyperbolic periodic attractor, the product attractor is not hyperbolic - it has $m$ Lyapunov exponents that are zero and in general we expect a very rich set of possible dynamics (including chaos) for arbitrarily small perturbations. This should gives a technique to prove the existence of stable weak chimeras in networks such as Figure~\ref{fig:modular_graphs}.

\section{Weak chimera states in non-modular networks}
\label{sec:nonmod}

For the modular networks considered in the previous section, the factorization into multistable modules enables one to understand weak chimeras as robust phenomena in such networks. It is also suggestive of the idea that chimeras are associate with ``spatial chaos'' - an exponential scaling of the number of attractors as the number of modules goes to infinity \cite{omelchenko-etal-2012}. Nonetheless, many of the chimeras that have hitherto been investigated in the literature do not have this modular structure. This section considers some six oscillator networks where there can be bifurcations to weak chimera states.

\subsection{Stable and neutral weak chimeras in six oscillator networks}

\begin{figure}%
\begin{center}
\includegraphics[width=9cm]{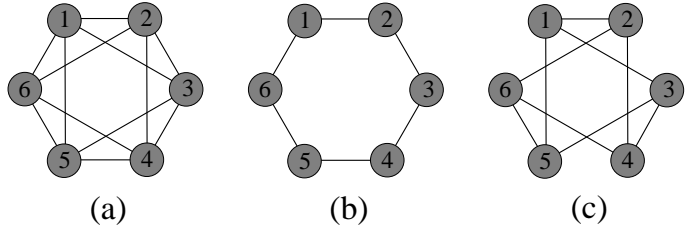}
\end{center}
\caption{(a) Six oscillators with nearest and next-nearest neighbour coupling. (b) Six oscillators with nearest neighbour coupling only. (c) Six oscillator system with three inputs to each oscillator; each of these networks has six indistinguishable oscillators and supports weak chimera states (see text for details).}%
\label{fig:six_networks}%
\end{figure}

Three non--global coupling structures of six indistinguishable oscillators are shown in Figure~\ref{fig:six_networks}.
For each of these networks and coupling (\ref{eq:hmmcoupling}) there can be attracting weak chimera states. For example, each of three systems has a stable weak chimera for $\alpha=1.6$, $r=-0.01$. 

Consider the network Figure~\ref{fig:six_networks}(a) governed by
\begin{equation}
\dot{\theta}_i = \omega + \sum_{|j-i|=1,2} g(\theta_i-\theta_j)
\label{eq:ringofsix}
\end{equation}
with coupling (\ref{eq:hmmcoupling}) and indices taken modulo $6$. 

Chimera states have been investigated in similar systems, for example by Maistrenko and co-workers; for example \cite{omelchenko-etal-2010,omelchenko-etal-2012} and transient chimeras have been found for coupling (\ref{eq:hmmcoupling}) with $r=0$ (Kuramoto-Sakaguchi), where the length of transient scales exponentially with the size of the system \cite{wolfrum-omelchenko-2011}.

\begin{figure}%
\begin{center}
\includegraphics[width=13cm]{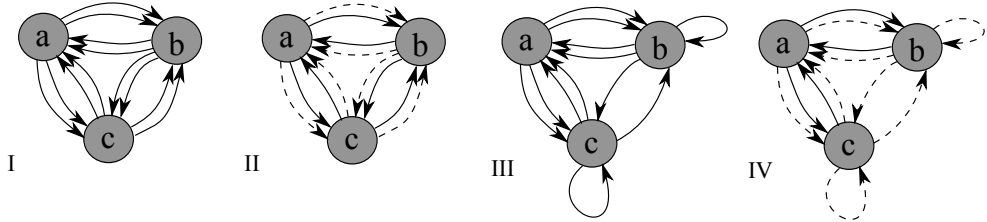}
\end{center}
\caption{Three-cell quotient networks of the network Figure~\ref{fig:six_networks}(a). The solid arrows denote an input to one cell from another while the dashed arrows indicate an input that includes a phase shift of the phase by $\pi$. Note that quotients I, II have symmetry $\D_3$ while III, IV have symmetry $\Z_2$.}%
\label{fig:ringofsix_quotients}%
\end{figure}

Table~\ref{tab:ringofsix_iso} summarises the invariant subspaces for (\ref{eq:ringofsix}), cf. \cite[Table~2]{ashwin-swift-1992}. In addition to symmetry-forced subspaces, the coupling structure means that there are a number of additional invariant subspaces associated with certain quotient networks; see Antoneli and Stewart \cite{antoneli-stewart-2006}. The three-cell quotients are illustrated in Figure~\ref{fig:ringofsix_quotients} (see also \cite{aguiar-dias-2007}).

\begin{table}%
$$
\begin{array}{lccc}
\mbox{Subspace} & \mbox{Typical point} & \mbox{Dim} & \mbox{Reduced system}\\
\Sigma & (\theta_1,\ldots,\theta_6) & & \\
\hline
\D_6 & (a,a,a,a,a,a) & 1 & \\
\D_6^- & (a,a+\pi,a,a+\pi,a,a+\pi) & 1 & \\
\Z_6^1 & (a,a+\zeta,a+2\zeta,a+3\zeta,a+4\zeta,a+5\zeta) & 1 & \\
\Z_6^2 & (a,a+2\zeta,a+4\zeta,a,a+2\zeta,a+4\zeta) & 1 & \\
\hline
\D_3 & (a,b,a,b,a,b) & 2 & \\
\Z_3 & (a,b,a+2\zeta,b+2\zeta,a+4\zeta,b+4\zeta) & 2 & \\
\hline
\D_2 & (a,b,a,a,b,a) & 2 & \\
\D_2^- & (a,b,a,a+\pi,b+\pi,a+\pi) & 2 & \\
\Z_2^1 & (a,b,c,a,b,c) & 3 & \mbox{I} \\
\Z_2^2 & (a,b,c,a+\pi,b+\pi,c+\pi) & 3 & \mbox{II} \\
\hline
A_0 & (a, b, c, a, d, e) & 5 \\
A_1 & (a,b,c,a,c,b) & 3 & \mbox{III} \\
A_2 & (a,b,b,a,c,c) & 3 & \mbox{III} \\
A_3 & (a,b,c,a+\pi,c+\pi,b+\pi) & 3 & \mbox{IV} \\
A_4 & (a,b,b+\pi,a+\pi,c+\pi,c) & 3 & \mbox{IV} \\
A_5 & (a,a+\pi,b,a,a+\pi,b) & 2 & \\
A_6 & (a,a+\pi,b,a,a+\pi,b+\pi) & 2 \\
A_7 & (a, a+\pi, b, a+\pi, a, b) & 2 
\end{array}
$$
\caption{Invariant subspaces for the six oscillator system Figure~\ref{fig:six_networks}(a) for $\zeta:=\pi/3$ and $a,b,c,d,e,f$ are arbitrary phases. The three-oscillator reduced systems are shown in Figure~\ref{fig:ringofsix_quotients}. The subspaces $A_i$ are not invariant due to symmetries; rather they are ``exotic balanced polydiagonals'' in the terminology of \cite{antoneli-stewart-2006} that are invariant due to the form of coupling in the system.}
\label{tab:ringofsix_iso}
\end{table}

\begin{figure}%
\begin{center}
\includegraphics[width=10cm]{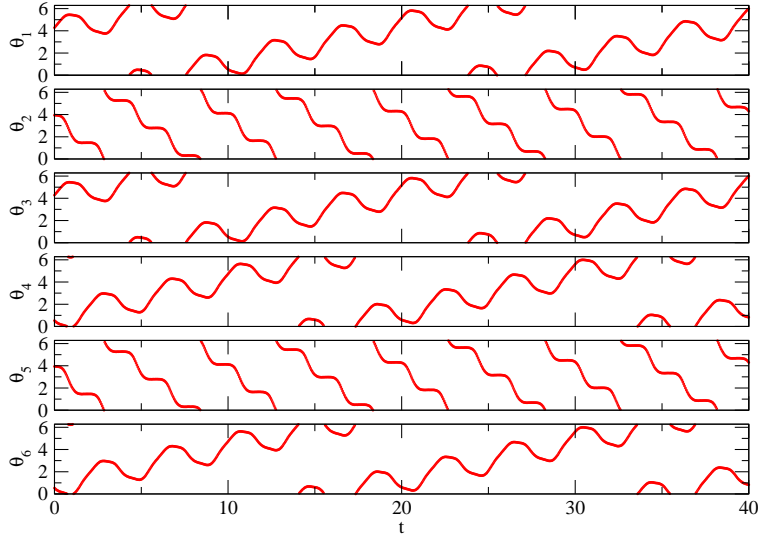}
\end{center}
\caption{A stable weak chimera state in the ring of six phase oscillators (\ref{eq:ringofsix},\ref{eq:hmmcoupling}) showing timeseries $\theta_i(t)$ for $\alpha=1.56$, $r=-0.1$. Observe that the frequency of the second and fifth oscillators clearly differ from the frequency of the others. This attractor coexists with in-phase synchrony.}%
\label{fig:ringofsixchimera}%
\end{figure}

There is an open set of parameters near $\alpha=1.56$, $r=-0.1$ where the system has stable weak chimeras that become marginally stable for $r\rightarrow 0$. Figure~\ref{fig:ringofsixchimera} illustrates such a solution that is in the invariant subspace $A_7\subset A_1$:
\begin{equation}
(\theta_1,\ldots,\theta_6) = (\phi_1,\phi_2,\phi_1,\phi_1+\pi,\phi_2,\phi_1+\pi).
\label{eq:synchsix}
\end{equation}
Interestingly, the same dynamics can be found within $A_1$ and $A_2$ as both have the quotient network III in Figure~\ref{fig:ringofsix_quotients}. Other invariant subspaces, for example the subspace $A_6$:
$$
(\theta_1,\theta_2,\theta_3,\theta_4,\theta_5,\theta_6)=
(\phi_1,\phi_1+\pi,\phi_2,\phi_1,\phi_1+\pi,\phi_2+\pi)
$$
has weak chimera solutions that are stable for $r=0$ and $\pi/2<\alpha<\pi$.

\subsection{Weak chimeras and bifurcations for the six-oscillator system}

\begin{figure}
\begin{center}
\includegraphics[width=16cm]{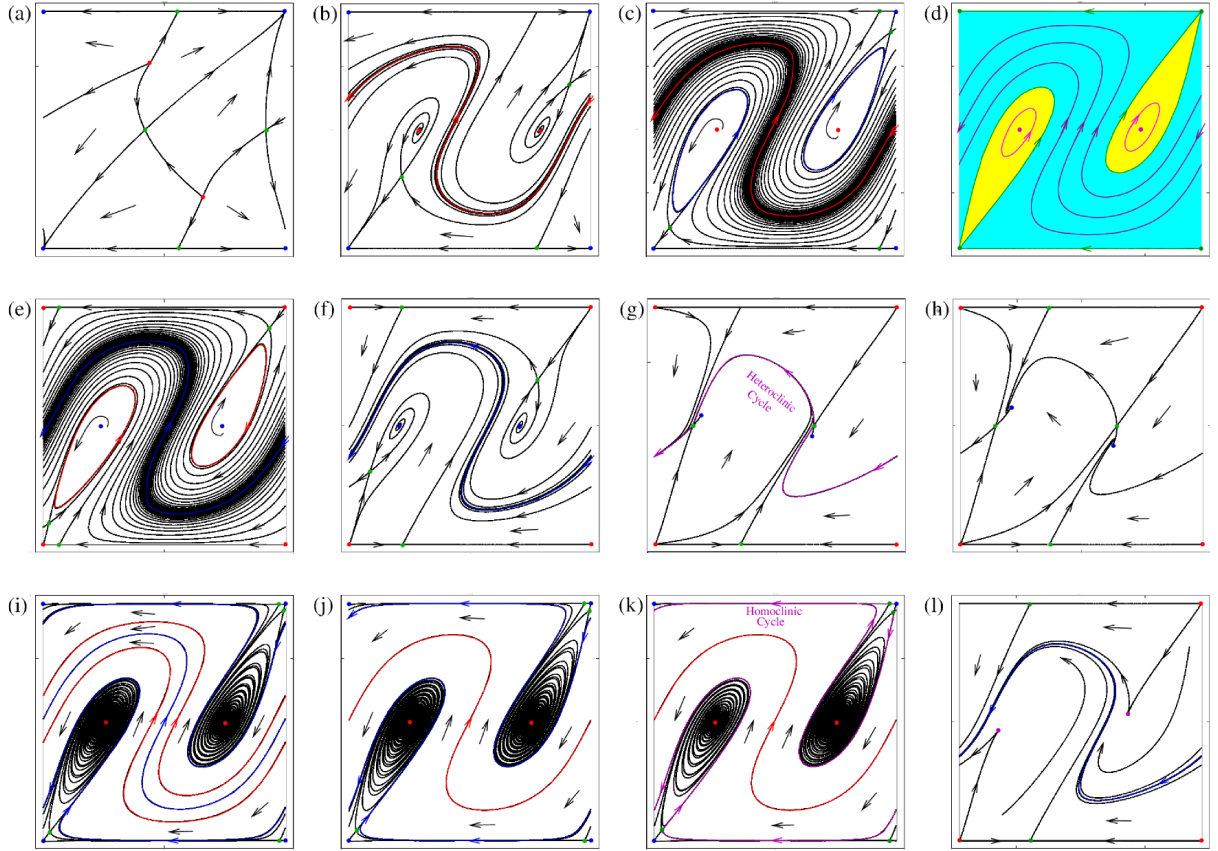}
\end{center}
\caption{\label{fig:ringofsixphaseplane}
Phase portraits for the reduced system (\ref{eq:ringofsixreduced_b},\ref{eq:hmmcoupling}) in the $\xi, \eta\in [0, 2\pi)$ plane. Red --- attractor, blue --- repellor, green --- saddle, magenta --- neutral, homo/heteroclinic cycle. The parameter values are as follows:
(a) $r=0$, $\alpha=0.5$, (b) $r=0$, $\alpha=1.3$, (c) $r=0$,
$\alpha=1.5$, (d) $r=0$, $\alpha=\pi/2$, (e) $r=0$, $\alpha=1.64$,
(f) $r=0$, $\alpha=1.84$, (g) $r=0$, $\alpha=2.16205$, (h) $r=0$,
$\alpha=2.22$, (i) $r=-0.01$, $\alpha=1.561$, (j) $r=-0.01$,
$\alpha=1.558$, (k) $r=-0.01$, $\alpha=1.5517$, (l) $r=-0.01$,
$\alpha=1.97794$. The periodic orbits that  wind around the $\xi$ direction of the torus are weak chimera states while the contractible periodic orbits are not weak chimeras; see text for more details.}
\end{figure}

\begin{figure}%
\begin{center}
\includegraphics[width=16cm,clip=]{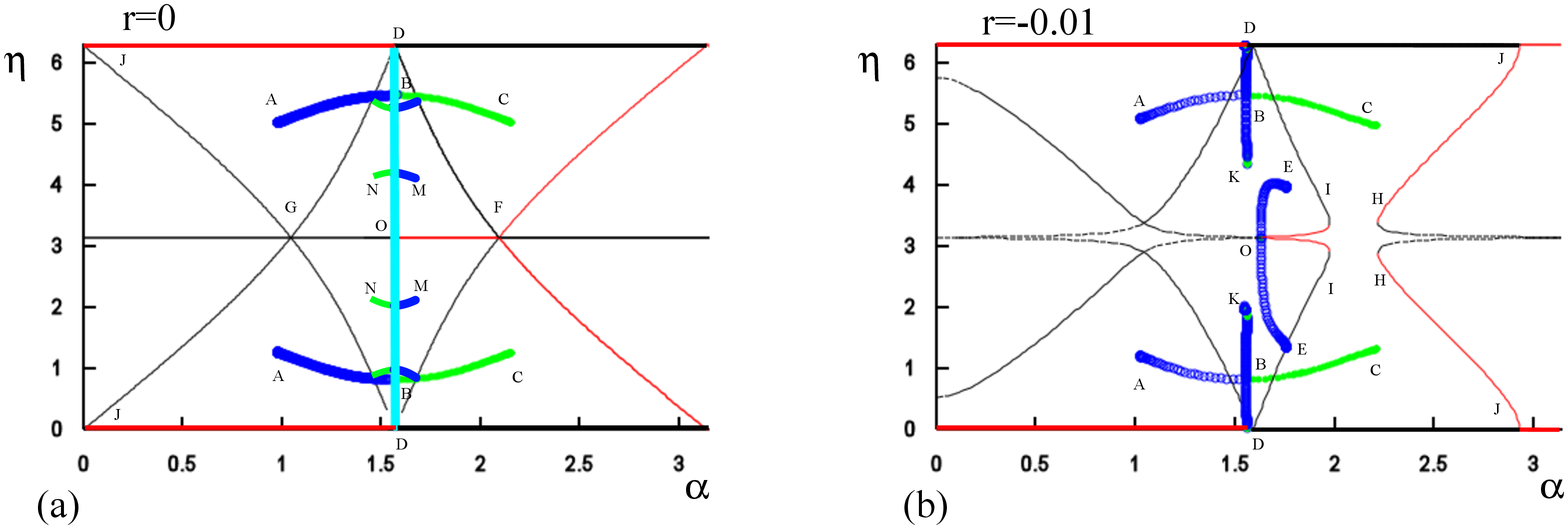}

~

\includegraphics[width=8cm,clip=]{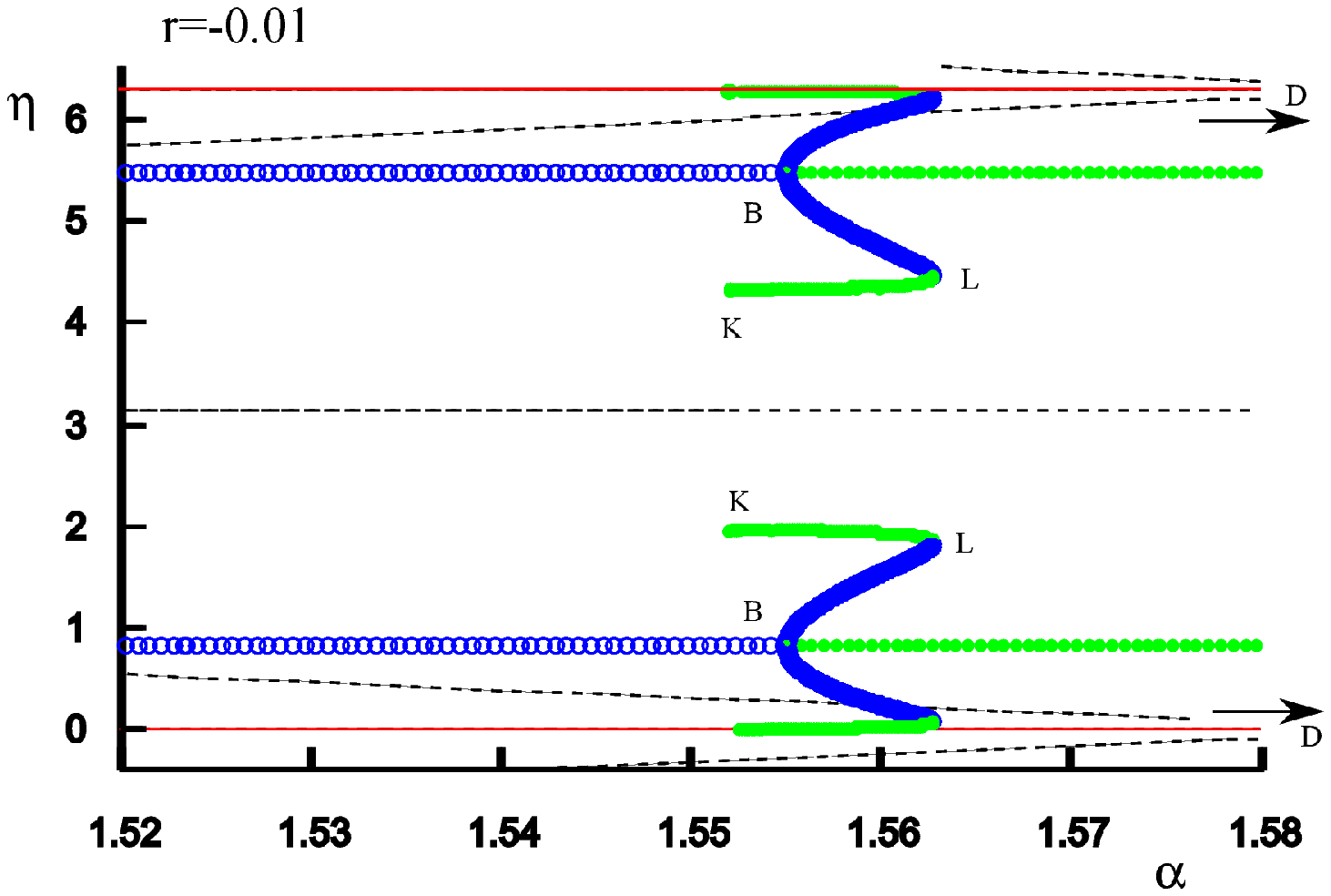}

\end{center}
\caption{Top: Bifurcation diagram for the reduced system (\ref{eq:ringofsixreduced_b},\ref{eq:hmmcoupling}) (a) for the Kuramoto-Sakaguchi case $r=0$ and (b) for $r=-0.01$. Red lines indicate stable equilibria, black are unstable equilibria. Green/blue/cyan lines indicate unstable/stable/neutral periodic orbits. Observe the bifurcations and non-generic ``vertical branches'' of periodic orbits for $\alpha=\pi/2$ that resolve into several generic branches of periodic orbits for $r\neq 0$ while $BC$ and $KL$ are branches of stable weak chimera states. Bottom: Close-up of some branches for $r=-0.01$; see text for more details.}%
\label{fig:ringofsixbifs}%
\end{figure}

We give a detailed (but not comprehensive) analysis of the dynamics of (\ref{eq:ringofsix}) in particular within $A_1$. Re-writing the system (\ref{eq:ringofsix}) in the subspace $A_1$ (\ref{eq:synchsix}) gives
\begin{equation}
\begin{split}
\dot{\phi}_1 &= \omega+2g(\phi_1-\phi_2)+2g(\phi_1-\phi_3)\\
\dot{\phi}_2 &= \omega+2g(\phi_2-\phi_1)+g(\phi_2-\phi_3)+g(0)\\
\dot{\phi}_3 &= \omega+2g(\phi_3-\phi_1)+g(\phi_3-\phi_2)+g(0)
\end{split}
\label{eq:ringofsixreduced}
\end{equation}
which corresponds to the three-oscillator quotient system III from Figure~\ref{fig:ringofsix_quotients}. Defining $\xi=\phi_1-\phi_3$, $\eta=\phi_2-\phi_3$ and $\xi-\eta=\phi_1-\phi_2$, the system (\ref{eq:ringofsixreduced}) can be written in terms of phase differences:
\begin{equation}
\begin{split}
\dot{\xi} & =2g(\xi-\eta)+2g(\xi)-2g(-\xi)-g(-\eta)-g(0)\\
\dot{\eta} & =2g(\eta-\xi)+g(\eta)-2g(-\xi)-g(-\eta).
\end{split}
\label{eq:ringofsixreduced_b}
\end{equation}
For coupling (\ref{eq:hmmcoupling}) and $\alpha=\pi/2$, $r=0$ this simplifies to
\begin{equation}
\begin{split}
\dot{\xi} & = -2\cos(\eta-\xi)+\cos \eta +1 \\
\dot{\eta} & = -2\cos(\eta-\xi)+2\cos \xi.
\end{split}
\label{eq:ringofsixreduced_c}
\end{equation}
The vector field (\ref{eq:ringofsixreduced_c}) has zero divergence - all equilibria are centres or saddles and any periodic orbit is neutrally stable. There is a ``band'' of neutrally stable weak chimera solutions that wind around $\xi$ and ``islands'' of neutrally stable periodic solutions that are not weak chimeras; see for example Figure~\ref{fig:ringofsixphaseplane}(d). Figure~\ref{fig:ringofsixbifs} shows the branches of equilibrium and periodic solutions on varying $\alpha$ for $r=0$. One can verify that there are stable weak chimera states within $A_1$ for $|\alpha-\pi/2|$ and small but non-zero $r$; these are connected via a homoclinic bifurcation to the branches in Figure~\ref{fig:ringofsixbifs}. Many of these are also stable transverse to $A_1$ though we do not compute these in detail.

Bifurcations for the system (\ref{eq:ringofsixreduced_b},\ref{eq:hmmcoupling}) were computed using XPPAUT \cite{ermentrout} and dstool \cite{dstool} and include the following.\footnote{Note that in order to path-follow weak chimeras, the trajectories are not closed curves in phase coordinates - instead one must embed into a higher dimensional system where they do close.} Lower case letters refer to Figure~\ref{fig:ringofsixphaseplane} while capital letters refer to Figure~\ref{fig:ringofsixbifs}. There is an Andronov-Hopf bifurcation for the contractible (non-chimera) cycle for $0<\alpha<\pi/2$ on increasing $r$ (for example, for $\alpha=1.5$, $r=0.011707$) and $B,O$. There is a homoclinic bifurcation of a non-chimera cycle at $N,M$; transition from (b) to (c) and (e) to (f). There is a saddle-connection for the weak chimera-cycle $A,C,E,K$ and (g), (k). There is a saddle-node bifurcation of two weak chimera-cycles at $L$ and a pitchfork bifurcation of three weak chimera cycles at $B$, with transition from (i) to (j). There is a saddle-node for the equilibria at $I,H$ and (l). There is a pitchfork of equilibria at (b) with $\alpha\approx 2.91$) which is degenerate for at $J,D$ $r=0$, $\alpha=0$ and $\alpha=\pi$.

For $r=0$ there is a line of degenerate bifurcations $D,B,O$ that are resolved into generic saddle node bifurcations $I,H$ on taking $r\neq 0$. For $r=0$, the only branch of stable weak chimeras $BC$ is for $\alpha>\pi/2$ while there can be multistability in the region $BL$ between in-phase, weak chimera and ``non-chimera'' periodic orbits for $r\neq 0$.

Finally, we note that the network Figure~\ref{fig:six_networks}(b) has attracting periodic weak chimera solutions in the invariant subspace $A_4$, 
while the network Figure~\ref{fig:six_networks}(c) has periodic weak chimera states belong to the invariant subspace $(a, b, c, c+\pi, b+\pi, a+\pi)$. 
The latter system also appears to have weak chimera states for the special case of Kuramoto--Sakaguchi coupling ($r=0$).

\section{Discussion}
\label{sec:discuss}

This paper proposes a definition of weak chimera state for indistinguishable networks of identical phase oscillator networks, based on nontrivial clustering of frequencies. Our definition makes only minimal restrictions on the dynamics and stability of a weak chimera; we find examples of quasiperiodic and heteroclinic chimeras but there is nothing to stop weak chimeras being chaotic in systems for higher $N$. Our definition is restrictive in that we only consider phase oscillators coupled through indistinguishable coupling, though this can be generalised e.g. to coupled chaotic oscillators with an observable whose average is different for different oscillators in an attractor of the network. It should be straightforward to extend the notion of indistinguishable oscillator network to coupled cell networks with one cell type \cite{golubitsky_stewart_06}.

We do not attempt here to characterize the behaviour of weak chimeras in the limit $N\rightarrow \infty$. The Antonsen-Ott ansatz \cite{ott-antonsen-2009} has been very successfully used to understand chimera states (for example, in \cite{laing-2009,laing-2010,omelchenko-2013,pikovsky-rosenblum-2011}), though the coupling we consider (\ref{eq:hmmcoupling}) only allows this ansatz to be applied in cases where there is the clear degeneracy $r=0$. As chimeras are associated with coexistence of ``coherent'' and ``incoherent'' clusters, a good definition of chimera will require a discussion of scaling properties of these cluster sizes, which we have not done here. These scaling of properties will need to be verified in families of networks rather than for individual networks.

Chimeras in larger systems are often observed to exhibit slow and random drift of the incoherent clusters, for example see \cite{wolfrum-omelchenko-2011}. This means that a stable chimera may have identical frequencies when computed over long enough timescales, unless the regions of different behaviour are ``pinned'' to fixed domains. We suggest that weak chimeras, while not stable in such a situation, will serve to organize the behaviour within the attractor.

Finally, our study suggests a reason why chimeras appear to be transients \cite{wolfrum-omelchenko-2011} for Kuramoto-Sakaguchi coupling in small systems of phase oscillators. Many of the weak chimeras for the four oscillator system (\ref{eq:model4osc},\ref{eq:hmmcoupling}) and the six oscillator system (\ref{eq:ringofsix},\ref{eq:hmmcoupling}) have degenerate stability for $r=0$. This means that transients near weak chimeras may have very long lifetimes. However, generic reductions of phase oscillator systems will have $r\neq 0$ \cite{kori-etal-2014} and non-degenerate stability.

\subsection*{Acknowledgments}

We thank the following for stimulating discussions in relation to this work: Danny Abrams, Ana Dias, Christian Kuehn, Carlo Laing, Yuri Maistrenko, Erik Martens, Mark Panaggio and Jan Sieber. We particularly thank the reviewers for making a number of comments that allowed us to improve the exposition in the paper, and we thank Yuri Maistrenko for suggesting the phrase ``weak chimera''.

\bibliographystyle{plain}

\def\cprime{$'$}

\end{document}